\documentstyle[aaspp4]{article}
\def\mg2{Mg$_2$}

\def\sec{\hbox{$^{\prime\prime}$}}
\def\fdg{\hbox{$.\!\!^\circ$}}
\def\fs{\hbox{$.\!\!^{\rm s}$}}
\def\farcm{\hbox{$.\mkern-4mu^\prime$}}
\def\farcs{\hbox{$.\!\!^{\prime\prime}$}}
\def\degd#1.#2{ #1\fdg#2 }                 
\def\mind#1.#2{ #1\farcm#2 }               
\def\secd#1.#2{ #1\farcs#2 }               
\def\hhh{\ifmmode {\rm ^h}              
         \else {${\rm ^h}$}
         \fi}
\def\sss{\ifmmode {\rm ^s}              
         \else {${\rm ^s}$}
         \fi}
\def\hms#1h#2m#3s{                      
                  \relax
                  \ifmmode #1^{\rm h}\,#2^{\rm m}\,#3^{\rm s}
                  \else \hbox{$#1^{\rm h}\,#2^{\rm m}\,#3^{\rm s}$}
                  \fi
                 }
\def\dms#1d#2m#3s{                      
                  \relax
                  #1\degr\,#2\arcmin\,#3\arcsec
                 }
\def\hmsd#1h#2m#3.#4s{                  
                      \relax
                      \ifmmode #1^{\rm h}\,#2^{\rm m}\,#3\fs#4
                      \else \hbox{$#1^{\rm h}\,#2^{\rm m}\,#3\fs#4$}
                      \fi
                     }
\def\dmsd#1d#2m#3.#4s{                  
                      \relax
                      #1\degr\,#2\arcmin\,#3\farcs#4
                     }

\begin{document}

\title{A subarcsecond resolution near-infrared study of Seyfert and
`normal'  galaxies: I.~Imaging data}

\author{Reynier F. Peletier$^{1}$}
\author{Johan H. Knapen$^{2,}$\altaffilmark{6}\altaffiltext{6}{Visiting 
Astronomer, Canada-France-Hawaii Telescope operated by the National Research 
Council of Canada, the Centre National de la Recherche Scientifique de 
France and the University of Hawaii}}
\author{Isaac Shlosman$^{3}$}
\author{D. P\'erez-Ram\'\i rez$^{2}$}
\author{D. Nadeau$^{4,6}$}
\author{R. Doyon$^{4,6}$}
\author{J.M. Rodriguez Espinosa$^{5}$}
\and
\author{A.M. P\'erez Garc\'\i a$^{5}$}

\affil{$^{1}$Department of Physics, 
University of Durham, South Road,
Durham, DH1 3LE, UK,
E-mail: R.F.Peletier@dur.ac.uk}

\affil{$^{2}$Department of Physical Sciences,
University of Hertfordshire,
Hatfield, Herts AL10 9AB, UK,
E-mail: knapen@star.herts.ac.uk}

\affil{$^{3}$ Department of Physics \& Astronomy,
University of Kentucky,
Lexington, KY 40506-0055, USA,
E-mail: shlosman@pa.uky.edu}

\affil{$^{4}$ Observatoire du Mont M\'egantic and 
D\'epartement de Physique,
Universit\'e de Montr\'eal,
C.P. 6128, Succursale Centre Ville,
Montr\'eal (Qu\'ebec), H3C 3J7 Canada,
E-mail: nadeau, doyon@astro.umontreal.ca}

\affil{$^{5}$ Instituto de Astrof\'\i sica de Canarias,
E-38200 La Laguna,
Tenerife, Spain,
E-mail: jre, apg@iac.es}

\journalid{Vol}{Journ. Date}
\articleid{start page}{end page}
\paperid{manuscript id}
\cpright{type}{year}
\ccc{code}
\lefthead{Peletier et al.}
\righthead{A near-infrared study of Seyfert galaxies: data}

\begin{abstract}

We present new high-resolution near-infrared observations in the $J,H$
and $K$ bands, obtained
to study the properties of Seyfert host galaxies. The dataset consists
of images in the three bands of practically the entire
CfA sample of Seyfert galaxies, and $K$-band images of a 
control sample of non-active, `normal', galaxies, matched to the Seyfert 
sample in the distribution of 
type and inclination. The spatial resolution  
and sampling of the new images is a factor 2
better than previously published $K$-band data. In this paper, we
present the data in the form of  profiles of
surface brightness and color, ellipticity and major axis 
position angle, as well as greyscale maps of surface brightness 
in $H$ or $K$ and both $J-H$ and $H-K$ 
colors. We  compare our surface brightness and 
color profiles with the literature, and find good agreement.

Our data are discussed in detail in three subsequent publications, where 
we analyze the morphologies of Seyfert and normal hosts, quantify the strength
of nonaxisymmetric features in disks and their relationship 
to nuclear activity,
address the question of bar fraction in Seyferts and normal galaxies, 
and analyze the color information in the framework of emission mechanisms 
in Seyfert~1s and 2s, and in non-active galaxies.
\end{abstract}


\section{Introduction}

By studying the relationship between active galactic nuclei (AGNs)
and their host galaxies we can learn about the processes that fuel
the central activity and the role the AGNs play in  galactic evolution. 
Because high spatial resolution is required for this purpose, the use 
of nearby samples is most promising.

In the last two decades a number of works have been dedicated to this 
issue. Early optical surveys (Adams 1977; Heckman 1978; Simkin, Su \& 
Schwarz 1980; Dahari 1984) studied the fraction of bars in Seyfert (Sy) 
galaxies, 
to find the link between Sy nuclear activity and non-axisymmetric 
distortions of the gravitational potentials of host galaxies by large-scale 
stellar bars and tidal interactions. This view was supported by theoretical
arguments that gravitational torques are able to remove the excess angular 
momentum from inflowing gas, leading to central or nuclear activity (e.g., 
reviews by Athanassoula 1994 and Phinney 1994). However, the early 
observational
results were questioned because the control samples were not matched
to the Sy sample in such important properties like morphological distribution (Balick
\& Heckman 1982; Fuentes-Williams \& Stocke 1988; Shlosman, Begelman
\& Frank 1990). Furthermore, it has been shown that a combination of dust
obscuration, stellar populations and inadequate spatial resolution can hide 
even a strong bar in the optical (e.g., Thronson et al. 1989; Block \& 
Wainscoat 1991; Spillar et al. 1992). 

It is clear that in the near-infrared (NIR), where one can look through 
the dust, more bars are found (McLeod \& Rieke 1995), but the fact that NIR 
arrays until recently covered small areas on the sky made galaxy 
classification in the NIR a very tedious job. Even very recent
studies (e.g. Moles, M\'arquez \& P\'erez 
1995; Ho, Fillipenko \& Sargent 1996) 
had to rely on the morphological classification from optical catalogs (e.g., 
the RC3, de Vaucouleurs et al. 1991). Only Mulchaey \& Regan (1997), using 
$K$-band observations with a field size of about 3$'$ $\times$ 3$'$, 
do not take this route, and conclude from NIR observations of Sy and control 
galaxies that the bar fractions are equal in these two classes.

Apart from studying the morphology, one can also make use of optical and NIR 
colors to investigate the nature of Sy galaxies. Traditionally, Sy
galaxies are classified into two categories, i.e., the Sy~1s 
with broad, and the Sy~2s with narrow  emission lines. 
The `unified model for AGNs' explains the 
observed differences between the two classes of Sy's in terms of 
orientation effects with respect to the observer (see Antonucci 1993 for a 
review). The model invokes an optically thick torus, which surrounds the
region responsible for the production of the broad lines (BLR) and is
oriented edge-on for Sy~2s, and face-on for Sy~1s. The main 
observational support for this scheme comes from
spectropolarimetry, showing scattered light from the BLR in Sy 2 galaxies,
and from striking bi-conical structures in line emission maps (e.g., 
NGC~5252, Tadhunter \& Tsvetanov 1989). 
NIR colors are generally well fitted 
by a combination of stellar populations and hot dust of about $1,000-1,300$~K 
(e.g., Glass \& Moorwoord 1985; Alonso-Herrero et al. 1998). 
Infrared color-color diagrams, like the $J-H$ vs. $H-K$ diagram, show that
the light in near the nuclei of many Seyfert galaxies cannot be reproduced
purely by starlight, even extincted by dust, because for a given $J-H$ 
the $H-K$ is just too red (e.g. Alonso-Herrero et al. 1998). 
$H-K$ colors larger than 1 are sometimes seen in Seyfert nuclei, 
which should be compared to the reddest stars, which have $H-K$ $\sim$ 0.45
(Bessell \& Brett 1988). It is thought that in such a case, a significant
fraction of the light comes from thermal radiation from dust heated by
a primary ultraviolet source, which could be the AGN or a site of star
formation (Rieke 1978). In the case of 3C273 Barvainis (1987) has shown 
that he could reproduce its spectrum, with a bump at 2 microns due to emission
by $\approx$ 700 M$_\odot$ of dust of $\approx$ 1000 K situated between
2 and 20 pc from the nucleus. Others (e.g. Alonso-Herrero et al. 1998) 
have shown that the position of Seyferts in the $J-H$ vs. $H-K$ diagram
is well fitted by such models. To see emission from hot dust, one has
to go to the $K$-band, since the $H$-band is generally barely affected.
Sy galaxies, in addition to 
hot dust, must also contain a large amount of cold dust, causing the optical
BLR to be completely obscured in Sy~2s.
This cold dust will extinguish and redden the colors of the
nucleus. High-resolution NIR color profiles will provide 
information about the extent of the region containing hot dust,
and in this way will constrain both the energy radiated by 
the central engine and the kinematics of the dusty torus.

This is the first in a series of 4 papers in which we study the morphology 
and photometry of Sy galaxies, and compare them with non-active (`normal') 
galaxies. For
this purpose, we have observed the  CfA sample, an unbiased sample 
of Sy~1s and 2s (Huchra \& Burg 1992) in the $J,H$ and $K$ bands
with a median seeing of \secd 0.7, 
i.e., more than a factor 2 better than in previous NIR surveys (McLeod 
\& Rieke 1995; Hunt et al. 1997; Mulchaey, Regan \& Kundu 
1997, Hunt et al. 1999). This data set is 
presented along with imaging of 
a control sample of non-active galaxies, matched to 
the CfA sample. To compensate for the fact that our field
size is only about 1$'$, we have combined our data in the outer parts of
the galaxies with ellipticity and position profiles obtained from the 
Palomar Sky Survey (Lasker et al. 1990).
Since in the outer parts of spirals the influence of extinction is
much less than near the center (e.g., Peletier et al. 1995), this will most
likely give results similar to those from large-size NIR arrays. Here 
(Paper~I), we describe the observations of the CfA and control samples, and
present the imaging data, accompanied by fits to the NIR \& POSS images. 
Knapen, Shlosman \& Peletier (1999, hereafter Paper~II)  
compare both samples to 
analyze the prevailing morphologies of Sy and normal hosts and their 
bar fractions.  Shlosman, Peletier \& Knapen (1999, Paper~III)  
address the issue of 
bar strengths in Sy and non-Sy galaxies. Peletier, Knapen \& Shlosman 
(1999, Paper~IV), 
analyze our photometric data in the framework of emission mechanisms
in Sy~1s and 2s, and in normal galaxies. 

This paper is organized as follows. In Section~2, we describe the 
samples and in Section~3
the observations. In Section~4, the data reduction steps are discussed, and 
in Section~5 we compare our observations with previous work in the
literature. A summary is given in Section 6. Finally, the 
appendix gives some notes on the individual objects.

\section{Samples}

The Sy sample considered here is the CfA sample (Huchra \& Burg 1992). It
is basically a spectroscopically selected blue-magnitude limited sample of 
active galaxies. We took all galaxies classified by Huchra \& Burg (1992)
as Sy~1s and 2s. The advantage of selecting Sy's in this
way is that it is relatively unbiased, and that one can make comparisons
between the properties of Sy's of types 1 and 2. The sample has been 
observed at many different wavelengths, and is therefore well-suited to study
the physical properties of AGNs. We excluded 3C~273 because of its large
redshift ($z=0.16$ vs. 0.07, for the second largest). No NIR observations
were taken of NGC~4395, since it is a dwarf galaxy, much fainter than the 
rest of the sample, with such a low surface brightness that we could barely
detect the nucleus. NGC~5347 was also not included, for observational
reasons. 
We did observe Mrk~789, which is part of the CfA sample, but was not included
in the study by McLeod \& Rieke (1995). Mrk~471, included by 
McLeod \& Rieke but not by
Huchra \& Burg, was not included. The sample is shown in Table~1.

We also observed a control sample
of 34 galaxies, selected from the RC3 to mimic the Sy sample
accurately in terms of morphological type  (including
barred/non-barred as derived from the RC3 classification) 
and ellipticity. In Paper~II, the detailed 
selection criteria of this sample are discussed, and the morphology of 
the galaxies in the two samples compared. The control sample is presented 
in Table~2. 


\section{Observations}

Observations were performed at the Observatorio del Roque de los Muchachos at 
La Palma for 2/3 of the sample, using WHIRCAM (Hughes, Roche \& Dhillon 1996)
at the 4.2m William Herschel Telescope (WHT), 
and at the 3.6m Canada France Hawaii Telescope (CFHT) at Hawaii 
for the remaining third, using the Montr\'eal NIR camera (MONICA; Nadeau
et al. 1994). At La Palma, the observations were made in September, 1995
and in April, 1996, whereas all CFHT  
observations were made in February, 1996.
WHIRCAM uses an InSb array with 256 $\times$ 256  pixels of \secd 0.{245} 
on the sky. MONICA uses a 256 $\times$ 256 HgCdTe array with a very similar 
pixelsize of \secd 0.{242}. To match the ellipticity distribution of the
Sy  and 
control samples, we added the galaxy NGC~6504 to the control sample.
This galaxy had been observed at UKIRT in 1994
using a 256 $\times$ 256 InSb array with pixelsize of \secd 0.{291}
(Peletier \& Balcells 1997).

The observing procedure consisted in taking a series of 4 dithered frames
of typically  60s each, flanked or followed by  
4 dithered sky frames of the same exposure time,  
taken at offsets of about 5 - 10$'$ from the object. After this, the same
procedure was repeated. A small linearity correction
was applied to the WHIRCAM data, determined from dome flatfields taken 
using various integration times. No linearity correction was needed
for the CFHT exposures. 

Care was taken that none of the exposures were
saturated, implying, for example, that the integration time per readout 
for NGC~4151 in $K$ was 0.5s. Our exposures for that galaxy, and to
a lesser extent for a few others, contain rather strong diffraction spikes,
due to the relative strength of the nuclear source. The CFHT exposures
suffer to a small extent from residual images. After observing 
a bright source, the next image may contain a faint pointsource at 
the position of the previous peak. The strength of these residual sources 
was 0.2\% of the initial source at most, and was only visible for galaxies
with very bright nuclei. This hysteresis effect
does not affect the data presented here. 

For the control sample, $K$-band  images of the same quality as those of
the Sy
sample were obtained. The control sample was observed during the same 
nights as the Sy sample, so the photometric and seeing conditions are the 
same for both samples. 

During the observations some high clouds were  present. Comparison
with the surface photometry of McLeod \& Rieke (1995) and aperture
photometry in the literature shows that the attenuation by clouds in $J$, 
$H$ and $K$ in general varied between 0 and 0.5 mag.

To estimate 
the quality of the images, we have measured the seeing for every galaxy
in each band using stars on the galaxy frame or on adjacent sky frames.
In some cases no stars were available, and seeing values have 
been taken from galaxy frames observed approximately at the same time at 
similar 
airmass values. The measured values for the FWHM of the images
have been given in Tables~3 and 4. They should be considered upper
limits for the seeing, given the fact that some images are almost 
critically sampled.
The seeing was generally excellent, and varied between \secd 0.5
and \secd 1.0. The reduced mosaics have an effective median seeing of about
\secd 0.7.

\section{Data Reduction}

The basic data reduction procedure was standard, and although different in
details for our WHT and CFHT data, can be summarized as follows.  After
subtracting an appropriate dark frame,  all sky frames belonging to a certain
object were median combined, and subtracted from each object frame. A flatfield
was created by taking the median of all normalized sky-frames taken during the
night. After flatfielding, a mosaic was made from all object frames by aligning
them on the centroid of the nucleus or a bright star. Bad pixels were  excluded
using a mask. This process is the same as, for example, described in Peletier
(1993). Further details of the CFHT data reduction procedure can be found in
P\'erez-Ram\'\i rez et al. (in preparation). The images are shown in Fig.~1.
The data presented in Figure~1 is tabulated in Table~5. Only tabulated are data
points for which the error in the $H$-band surface brightness is less than 1
mag. 

During each run, the position angle (from N to E) of the camera was
measured by comparing the positions of stars on our frames with images
of the POSS Digital Sky Survey (ref.). In the same way our pixel size
was determined to be \secd 0.{245} $\pm$ 0.005\% for the WHIRCAM images,
and \secd 0.{242} $\pm$ 0.005\% for the MONICA frames. These numbers
were used to determine the final radial morphological profiles.

To study the morphology of the galaxies in the Sy and Control samples,
we fitted ellipses to the images using {\tt galphot} (see J\o rgensen,
Franx \& Kjaergaard 1992).
We first removed foreground stars by hand, and did not 
consider them in the fit. The center position was allowed to change freely as
a function of radius.  After ellipse fitting,
we determined the residual background sky value for each frame, and
subtracted it from the profile. This was done by taking a median value 
of the signal in the outer parts of the frame, which were thought to be
the least affected by galaxy emission. For large galaxies a better 
estimate was made by assuming that the logarithmic slope of the surface brightness
profile remained constant outside the frame. Although this might lead to
quite large errors, the contribution of most galaxies in the outer parts of
the frames is negligible. As an example, for all but the brightest
galaxies, the surface brightness in $H$ at the edge of the frame 
is lower than 21 mag\,arcsec$^{-2}$, which is 0.17\% of a typical
sky surface brightness of 14 mag\,arcsec$^{-2}$. 
Background sky determination errorbars are 
included in Fig.~1. 1$\sigma$ errorbars correspond to $\mu_H$ =
21.5\,mag\,arcsec$^{-2}$, 
$\mu_J$=21.2\,mag\,arcsec$^{-2}$ and $\mu_K$=20.5\,mag\,arcsec$^{-2}$, 
which corresponds to about 0.1\% of the sky
background in $H$ and $K$, and 0.3\%  in $J$. 
The dotted line corresponds to a radius of 1 seeing FWHM, outside which 
the effects of seeing on the ellipticity and position angle profile
in general are small. However, some of our
La Palma data suffer from an elongated PSF, as a result of bad tracking
of the telescope. In such a case the galaxy images in the inner regions 
are elongated towards PA 90 degrees, and up to a maximum of 2-3$''$
these profiles can not be trusted. The effect is present in all
frames observed at La Palma, and is worst for the most elongated
galaxies.

In order to increase the field size,
we also took the images of the digitized Palomar
Observatory Sky Survey (POSS; E-plates), 
and fitted ellipses to them in the same 
way as described above. Although most of these images are saturated in the 
central regions, they extend further out than would equivalent NIR images,
since the POSS images are deep
$R$-band images, and galaxies generally become
less dusty when going outward, which implies that they are bluer.
One of the reasons that our observations have been
taken in the NIR is the fact that spiral galaxies, and especially
Sy's are dusty in general (e.g., Valentijn 1990; Beckman et al. 
1996: Alonso-Herrero et al. 1998). However, the amount of extinction 
generally decreases when going outward. Peletier et al. (1995) showed that
the extinction at 3~NIR disk scale lengths in $B$ is
smaller than 0.5 mag, in general.
Our NIR data cover in all cases the inner three scale lengths of
our galaxies. Further out, where we have to use the red POSS data, the
influence of extinction can be neglected. Agreement between
NIR and POSS datasets in the overlap region is generally good. 
The POSS fits have been plotted along with
the NIR fits in Fig.~1. In the inner regions no POSS-fit data are
shown when those images are saturated.

The observations were all done using partially-clouded weather
conditions, with high cirrus clouds present on many of the observing
nights. 
In particular
the 11th and 12th of September, 1995 suffered from clouds. Many of the 
galaxies observed on those nights were reobserved on the 13th. The weather
conditions in February and April, 1996 were better, with less cirrus. Since 
on April 28, 1996 the seeing was bad, all observations taken on that 
day were repeated later. Because of the clouds, the Sy data were all 
photometrically calibrated using aperture photometry in the 
literature. This is not an ideal 
situation, since many of these AGNs are variable (e.g., Peterson et al. 1998).
We could however find aperture photometry in $J$, $H$ and $K$ for all 
but 2 of the galaxies, Mark~590 and UGC~6100, which we calibrated 
using standard stars observed during the night. Most of the literature
aperture photometry was obtained from Edelson, Malkan \& Rieke  (1987; 
26 galaxies).
Other comparison sources included Spinoglio et al. (1995; 11), Rieke
(1978; 4), Rudy et al. (1982; 2) and McAlary (1982; 1). As a check, 
we compared our $K$-band data with the surface photometry of
McLeod \& Rieke (1995). In general the agreement was reasonable. For
NGC~5940, however, using photometry of Edelson et al. (1987) 
there was more than a magnitude difference
between the two calibrations, and for that reason we applied our 
standard star calibration. 
We claim that the absolute accuracy of our 
photometry is about 0.3 mag (1$\sigma$ RMS), based on the comparison 
with McLeod \& Rieke (1995). We refer the reader to Section~5 for 
further details on comparison with the literature.

For the control sample, we used standard stars. On each night 
in periods without much cloud cover many standard 
stars were observed from the list of Casali \& Hawarden 
(see Hughes et al. 1996) and Carter \& Meadows (1995). 
For each run, the 
same photometric calibration was used, and no corrections were applied
for atmospheric and galactic extinction, nor were any color terms applied.
Comparison with the Seyfert galaxy data, which was photometrically calibrated
using aperture photometry in the literature, shows that the photometric
accuracy of the control sample is about 0.3 -- 0.5 mag. The importance
of this number is marginal however, since no color information is provided
for the control sample. 

After the photometric calibration, color index maps and profiles were made
of the Sy's. To create color maps,  frames in two
bands were first sky-subtracted, after which the frame with the best
seeing was convolved with a gaussian to match the circular gaussian 
seeing FWHM of both galaxies. The frames were then shifted to the same
center, divided by each other, and calibrated. 
The resulting color maps are  shown in Fig.~ 1. In a few
cases  some effects of different PSFs are seen in the center, 
since in general the PSF on the WHT is slightly elliptical due to
telescope movements during the observations, an effect noticed 
especially when the seeing is good. For a typical seeing of \secd 0.7,
the motion of the telescope gives a round object an ellipticity of 0.2.
After fitting ellipses to the $H$-band images, we
fitted ellipses to the $J$ and $K$ images, keeping the shape 
of the isophotes constant. In this way radial color profiles could
be determined. In Fig.~1 we show $J-H$ and $H-K$ color profiles.
In the figure, dotted lines indicate a radius corresponding to
1 seeing FWHM. Simulations have shown (e.g., 
Franx, Illingworth \& Heckman 
1989; Peletier et al. 1990) that generally differences in seeing will not 
produce errors larger than 0.05 mag at radii larger than indicated by 
this line. 

\section{Comparison with the Literature}

To check the quality of our photometry, we have made a number of
comparisons with literature data. The surface brightness profiles
in $K$ were compared to
the NIR array observations of McLeod \& Rieke (1995) in $K$. The comparison
is shown in Fig.~2 and is good in general. In the central regions, our 
data tend to be brighter, due to our better seeing. In individual cases, 
however, this effect 
can be caused by variability in the luminosity of the AGN itself. 
Peterson et al. (1998), using an aperture of $5\sec ~\times$ \secd 7.6 
characteristically find a variability in the optical continuum flux around 
5,000 \AA\ of 10-30\% (RMS). It appears that in the NIR, the amplitude of the 
variability
is similar to that in the visual region. The origin of variability
in the NIR apparently is heating of a dusty circumnuclear
region by a variable nuclear source (Glass 1998 and refs. therein). 
The data in most cases show 
that this circumnuclear region has a temperature of about $1,000-1,300$~K
(Glass \& Moorwood 1985; Alonso-Herrero et al. 1998), which
implies that the variability in $K$ is somewhat stronger than in $H$ and 
$J$. 

Somewhat further away from the center, between about 3 and 10$''$ in
radius, the 
difference between our results and those of McLeod \& Rieke (1995) is 
generally constant as a function of radius. 
This means that the same surface brightness profile slope is measured,
and hence the difference in surface brightness is caused by
errors in the photometric calibration.
The average offset between McLeod \& Rieke and our results is 0.19 mag, 
with an RMS scatter of 0.28 mag, implying that their data are slightly 
brighter. The agreement between
the ellipticity profiles is good. Only near the center 
the profiles of McLeod \& Rieke typically are rounder, generally
due to seeing effects. For a few galaxies, we do not present an ellipticity
profile. These galaxies are severely distorted,
making it impossible to make unambiguous ellipse fits.

 
We compared the color profiles to the data of Hunt et al. (1997), where
we have 15 galaxies in common. Apart from occasional differences
of a few tenths of a magnitude in the color zeropoint level, the $J-H$
and $H-K$ color profiles are very similar. Red features near the nucleus
are always confirmed in the profiles of Hunt et al. (1997), 
except in a few cases, where the difference in seeing FWHM clearly affects
their profiles. 

Recently Hunt et al. (1999) published an atlas with surface photometry
of 90 Seyferts of type 1 and 2. For most of their galaxies they present
infrared photometry in at least one band, while for many galaxies also
optical photometry is presented. The median seeing of their infrared data
is $\approx$ 1.5\sec. Their ellipticity and position angle profiles are
in reasonable agreement with the data published here. Comparing simulated
aperture photometry for the 16 galaxies in common in apertures of 
10\sec, 20\sec and 30\sec with the
values presented in their Table 2, we find
that our data are on the average resp. 0.09, 0.10 and 0.11 mag fainter, with a 
scatter of resp. 0.33, 0.31 and 0.31 mag, giving the same conclusion
as our comparison with McLeod \& Rieke (1995).

\section{Summary}

In this paper, we have presented a subarcsec resolution $J,H$ and $K$ 
imaging dataset to study the properties of the host galaxies of Sy nuclei.
It consist of images in the three bands of all but 2 galaxies of the 
CfA sample of Sy galaxies, a reasonably unbiased sample of nearby
Sy's~1 and 2, and $K$-band images of a control sample, matched to
the Sy sample, and observed to study morphological properties.
The seeing and sampling is about a factor 2 better than
data published previously by 
McLeod \& Rieke (1995). Here  we provide for surface brightness
and color profiles, ellipticity and major axis position angle 
profiles, including fits to the digital POSS survey, and greyscale 
maps of surface brightness and both $J-H$ and $H-K$ colors. We have 
compared the color profiles with Hunt et al. (1997) for 15 galaxies
in common, and the agreement is excellent. The comparison of the 
$K$-band surface brightness profiles with McLeod \& Rieke (1995) also
shows good agreement.

The data presented in this paper is analyzed in three companion papers.
Knapen et al. (1999), Shlosman et al. (1999) and Peletier et al.  
(1999), discuss bar properties and morphologies of Sy and normal hosts,
address the issue of bar fraction in Sy's and
non-Sy's, and focus
on the color information presented here, in the framework of 
emission mechanisms in Sy~1s and 2s, and in normal galaxies.

\acknowledgements

This paper is based in part on observations obtained at the William
Herschel Telescope, which is operated at La Palma by the Isaac Newton Group
in the Spanish Observatorio del Roque de los Muchachos of the Instituto 
de Astrof\'\i sica de Canarias.
We thank Dr. Kim McLeod and Dr. Leslie Hunt for sending us their data
tables.  This research has made use of the NASA/IPAC Extragalactic
Database (NED) which is operated by the Jet Propulsion Laboratory,
California Institute of Technology, under contract with the National
Aeronautics and Space Administration.  Supported in part by NASA grants
NAGW-3841 and WKU-522762-98-06, and HST AR-07982.01-96A (IS).  We used
images from the Digitized Sky Survey, which was produced at the STScI
under U. S. Gov. grant NAG W-2166, and is based on photographic data
obtained using Oschin Schmidt Telescope, operated by the California
Institue of Technology and Palomar Observatory. We made use of 
observations made with the NASA/ESA HST, obtained from the data
archive at STScI.  STScI is operated by AURA, Inc.  under NASA
contract NAS 5-26555.

\appendix

\section{Notes on the Individual Objects}

In this appendix, we indicate specific features relevant to our study.
We do not pretend to give a complete review of each object, or even to
give up-to-date references for them. Most of the observations given here
are directly based on our own images. In those cases where HST imaging
is available from the HST archive, we have checked those images for
relevant details, and mention them where appropriate, again without
pretending completeness. For some galaxies we have nothing to add to the
knowledge already existing in the literature, we did however list those
galaxies below in order to give the complete listing of our sample
galaxies. 

\subsection{CfA Sample}

\begin{itemize}
\item{Mrk 334: Has some faint outer tails. HST: disturbed}
\item{Mrk 335: Compact appearance, except for a jet-like feature to the NW.
HST: quasar-like}
\item{UGC 524: Face-on interacting spiral with outer arms.}
\item{1 Zw 1: Irregular, point-source dominated object.}
\item{Mrk 993: --}
\item{Mrk 573: Inner structure (inside 3$''$) possibly perpendicular
to intermediate region ($<$ 40$''$), which in turn is perpendicular to 
an outer envelope, as visible on the POSS image.}
\item{UGC 1395: --}
\item{Mrk 590: --}
\item{NGC 1068: Nearby AGN with strong bar, lens and tightly wound spiral arms.}
\item{NGC 1144: --}
\item{Mrk 1243: Inner bar perpendicular to rest of galaxy}
\item{NGC 3227: Large nearby galaxy, closely interacting with NGC~3226.}
\item{NGC 3362: Many spiral arms.}
\item{UGC 6100: --}
\item{NGC 3516: Inner bar (at appr. 10$''$) in an outer envelope, very well
visible on the POSS image.}
\item{Mrk 744: Closely interacting with NGC~3788.}
\item{NGC 3982: Circumnuclear SF}
\item{NGC 4051: Strong inner bar; irregular spiral arms, and fairly
round in the outer parts.}
\item{NGC 4151: Round bulge, with a strong outer bar. The very strong, red, 
central source makes the spikes of the telescope very well visible in 
the color maps. Some instrumental effects (ripples and low counts on 
one vertical semi-axis through the center) are visible as well.}
\item{NGC 4235: --}
\item{Mrk 766: HST: bar}
\item{Mrk 205: Very close to NGC 4319, but with very different redshifts
(0.07 for Mrk 205 vs. 0.006 for NGC 4319).}
\item{NGC 4388: Dusty edge-on galaxy}
\item{NGC 4395: Nearby, low surface brightness dwarf galaxy, not observed 
in the NIR}
\item{Mrk 231: Irregular appearance. Faint tails in the outer parts seen
on the POSS image and in de Robertis et al (1998). }
\item{NGC 5033: --}
\item{Mrk 789: Strongly interacting pair.}
\item{UGC 8621: Three-armed spiral.}
\item{NGC 5252: Featureless galaxy with slightly rounder outer regions. HST:
polar ring}
\item{Mrk 266: Closely interacting pair}
\item{Mrk 270: Regular appearance. Has a small inner bar visible in the NIR.}
\item{NGC 5273: --}
\item{Mrk 461: --}
\item{NGC 5347: Strongly barred galaxy, not observed in this paper.}
\item{Mrk 279: The galaxy has a large, resolved red inner region, only 
visible in $H-K$}
\item{NGC 5548: Peculiar outer shells.}
\item{NGC 5674: Face-on galaxy with a peculiar outer ring. The HST images 
show an inner bar perpendicular to the main bar.}
\item{Mrk 817: HST: bar}
\item{Mrk 686: --}
\item{Mrk 841: Compact, pointsource do\-mi\-na\-ted source}
\item{NGC 5929: Closely interacting with NGC~5930.}
\item{NGC 5940: Strong bar in face-on galaxy. HST: bar}
\item{NGC 6104: Strong-barred, interacting system with shells. HST: bar}
\item{UGC 12138: Barred galaxy getting rounder in outer parts. HST: bar}
\item{Mrk 335: Compact appearance, except for a jet-like feature to the NW. HST: quasar-like}
\item{NGC 7469: Peculiar isophotes, interacting galaxy}
\item{Mrk 530: -- }
\item{Mrk 533: Face-on spiral with a strong bar and a companion.}
\item{NGC 7682: Strong bar in face-on galaxy. HST: bar}
\end{itemize}

\subsection{Control Sample}

\begin{itemize}
\item{NGC 1093: Strongly-barred}
\item{UGC 3247: --}
\item{UGC 3407: Strongly barred galaxy, almost round in outer parts}
\item{UGC 3463: --}
\item{UGC 3536: Apparently edge-on, but outer regions getting round.}
\item{UGC 3576: Outer regions much rounder than part visible on the
NIR image.}
\item{UGC 3592: Strongly-barred, with bright star nearby}
\item{NGC 2347: --}
\item{UGC 3789: Strong-barred round, face-on galaxy}
\item{NGC 2365: --}
\item{UGC 3850: Galaxy has a large, round outer ring}
\item{NGC 2431: Inner bar perpendicular to rest of galaxy}
\item{NGC 2460: --}
\item{NGC 2487: Strongly-barred round galaxy}
\item{NGC 2599: Large position angle twist}
\item{NGC 2855: --}
\item{NGC 3066: Strongly-barred round galaxy}
\item{NGC 3188: Strongly-barred round galaxy}
\item{NGC 3455: --}
\item{NGC 4146: --}
\item{NGC 4369: --}
\item{NGC 4956: --}
\item{NGC 4966: --}
\item{NGC 5434: Close companion of UGC 8967}
\item{NGC 5534: Very irregular appearance of an interacting galaxy}
\item{NGC 5832: Strong PA twist in the outer parts.}
\item{NGC 5869: Ellipticity profile almost constantly rising}
\item{UGC 9965: Inner bar in round galaxy}
\item{NGC 5992: Face-on galaxy with wide spiral arms. Interacting
with NGC~5993}
\item{NGC 6085: --}
\item{NGC 6278: --}
\item{NGC 6504: edge-on}
\item{NGC 6635: --}
\item{NGC 6922: Peculiar spiral arms}
\end{itemize}

\vfill\eject

\vfill
{\small
\begin{table}[h]
\begin{center}
\begin{tabular}{llcrrlrrrcrrc}
\\
\\
\hline
\hline
Galaxy & Sy &  Type & \multicolumn{3}{c}{RA} & \multicolumn{3}{c}{Dec} &
z & M$_K$ & M$_B$ & MGT\\ 
\hline
Mrk 334   & 1.8 &  .P.....  &  00&03&09.622& +21&57&36.56  &0.0220  &  -23.8 &
-20.1 & RS1\\
Mrk 335   & 1 &    .P.....  &  00&06&19.519& +20&12&10.49  &0.0259  &  -24.2 &
-20.7 & SS1\\
UGC 524   & 1 &    PSBS3..  &  00&51&35.010& +29&24&04.53  &0.0359  &  -24.7 &
-21.3 & \\
1 Zw 1    & 1 &    .S?....  &  00&53&34.940& +12&41&36.20  &0.0604  &  -24.4 &
-22.4 & \\
Mrk 993   & 2 &    .S..1..  &  01&25&31.464& +32&08&11.43  &0.0154  &  -23.6 &
-19.4 & RS2\\
Mrk 573   & 2 &    RLXT+*.  &  01&43&57.802& +02&20&59.65  &0.0173  &  -23.8 &
-19.5 & RS2\\
UGC 1395  & 1.9 &  .SAT3..  &  01&55&22.039& +06&36&42.65  &0.0174  &  -23.7 &
-19.9 & \\
Mrk 590   & 1.2 &  .SAS1*.  &  02&14&33.562& -00&46&00.09  &0.0263  &  -24.8 &
-20.5 & SS1\\
NGC 1068  & 2 &    RSAT3..  &  02&42&40.711& -00&00&47.81  &0.0037  &  -24.7 &
-19.3 & \\
NGC 1144  & 2 &    .RING.B  &  02&55&12.196& -00&11&00.81  &0.0288  &  -25.5 &
-19.8 & RS2\\
Mrk 1243  & 1 &    .S..1..  &  09&59&55.835& +13&02&37.76  &0.0353  &  -24.1 &
-19.9 & \\
NGC 3227  & 1.5 &  .SXS1P.  &  10&23&30.589& +19&51&53.99  &0.0038  &  -23.8 &
-16.5 & SS1\\
NGC 3362  & 2 &    .SX.5..  &  10&44&51.716& +06&35&48.24  &0.0227  &  -24.7 &
-21.6 & RS2\\
UGC 6100  & 2 &     .S..1?. &  11&01&33.999& +45&39&14.16  &0.0291  &  -24.2 &
-20.9 & RS2\\
NGC 3516  & 1.5 &  RLBS0*.  &  11&06&47.490& +72&34&06.88 &0.0085  &  -23.6 &
-19.3 & SS1\\
Mrk 744   & 1.8 &  .SXT1P.  &  11&39&42.551& +31&54&33.43 &0.0091  &  -22.3 &
-18.1 & SS1\\
NGC 3982  & 2 &    .SXR3*.  &  11&56&28.102& +55&07&30.58 &0.003   &  -21.9 &
-19.1 & RS2\\
NGC 4051  & 1 &    .SXT4..  &  12&03&09.614& +44&31&52.80 &0.0022  &  -23.3 &
-16.2 & SS1\\
NGC 4151  & 1.5 &  PSXT2*.  &  12&10&32.579& +39&24&20.63 &0.0030  &  -23.9 &
-18.3 & \\
NGC 4235  & 1 &    .SAS1./  &  12&17&09.904& +07&11&29.08 &0.0077  &  -23.8 &
-19.0 & RS1\\
Mrk 766   & 1.5 &  PSBS1*.  &  12&18&26.509& +29&48&46.34 &0.0128  &  -23.3 &
-19.1 & SS1\\
Mrk 205   & 1 &    .P.....  &  12&21&44.120& +75&18&38.25 &0.070   &  -24.7 &
-23.5 & \\
NGC 4388  & 2 &    .SAS3*/  &  12&25&46.701& +12&39&40.92 &0.008   &  -22.5 &
-16.9 & \\
NGC 4395  & 1.8 &  .SAS9*.  &  12&25&48.918& +33&32&48.43 &0.0011  &  -21.6 &
-14.3 & \\
Mrk 231   & 1 &    .SAT5\$P &  12&56&14.2344& +56&52&25.236 &0.0410  &  -26.3 &
-21.3 & \\
NGC 5033  & 1.9 &  .SAS5..  &  13&13&27.526& +36&35&38.08 &0.0030  &  -24.4 &
-18.3 & \\
Mrk 789   & 1 &    .P.....  &  13&32&23.984& +11&06&20.19 &0.032   &  -24.0 &
-20.7 & \\
UGC 8621  & 1.8 &  .S?....  &  13&37&39.870& +39&09&16.99 &0.0201  &  -23.8 &
-20.1 & \\
NGC 5252  & 1.9 &  .L.....  &  13&38&15.963& +04&32&33.29 &0.0231  &  -24.6 &
-19.5 & RS1\\
Mrk 266   & 2 &    .P.....  &  13&38&17.69&  +48&16&33.9 &0.0275  &  -23.6 &
-20.9 & RS2\\
Mrk 270   & 2 &    .L...?.  &  13&41&05.759& +67&40&20.32 &0.0090  &  -22.7 &
-17.7 & RS2\\
NGC 5273  & 1.9 &  .LAS0..  &  13&42&08.338& +35&39&15.17 &0.0036  &  -21.8 &
-16.6 & \\
Mrk 461   & 2 &    .S.....  &  13&47&17.745& +34&08&55.34 &0.016   &  -23.1 &
-19.3 & \\
NGC 5347  & 2 &    PSBT2..  &  13&53&17.834& +33&29&26.98 &0.0304  &  -22.6 &
-18.9 & RS2\\
Mrk 279   & 1 &    .L.....  &  13&53&03.447& +69&18&29.57 &0.036   &  -24.5 &
-20.2 & \\
NGC 5548  & 1.5 &  PSAS0..  &  14&17&59.534& +25&08&12.44 &0.0166  &  -24.1 &
-19.7 & SS1\\
NGC 5674  & 1.9 &  .SX.5..  &  14&33&52.243& +05&27&29.65 &0.0248  &  -24.7 &
-21.2 & RS1\\
Mrk 817   & 1.5 &  .S?....  &  14&36&22.068& +58&47&39.38 &0.0314  &  -24.7 &
-21.1 & SS1\\
Mrk 686   & 2 &    .SB.3..  &  14&37&22.123& +36&34&04.11 &0.0122  &  -23.7 &
-19.1 & RS2\\
Mrk 841   & 1 &    .......  &  15&04&01.201& +10&26&16.15 &0.0364  &  -24.7 &
-20.5 & \\
NGC 5929  & 2 &    .S..2*P  &  15&26&06.161& +41&40&14.40 &0.0083  &  -22.0 &
-18.1 & RS2\\
NGC 5940  & 1 &    .SB.2..  &  15&31&18.070& +07&27&27.91 &0.0339  &  -24.8 &
-20.0 & US1\\
NGC 6104  & 1.5 &  .S?....  &  16&16&30.686& +35&42&29.00 &0.0280  &  -24.3 &
-20.9 & RS1\\
UGC 12138 & 1.8 &  .SB.1..  &  22&40&17.048& +08&03&14.09 &0.0250  &  -23.6 &
-20.6 & US1\\
NGC 7469  & 1 &    PSXT1..  &  23&03&15.623& +08&52&26.39 &0.0160  &  -25.0 &
-20.3 & SS1\\
Mrk 530   & 1.5 &  .SAT3*P  &  23&18&56.617& +00&14&38.23 &0.0290  &  -25.3 &
-20.5 & SS1 \\
Mrk 533   & 2 &    .SAR4P.  &  23&27&56.724& +08&46&44.53 &0.0289  &  -25.2 &
-20.8 & SS2\\
NGC 7682  & 2 &    .SBR2..  &  23&29&03.928& +03&32&00.00 &0.0170  &  -23.9 &
-19.6 & RS2\\
\hline
\end{tabular}
\end{center}
\caption{Properties of our Sy sample galaxies. Galaxy name (col. 1), 
Sy Type (from Huchra \& Burg 1992, col. 2),
morphological classification (from RC3, col. 3), RA and Dec (J2000, 
from NASA Extragalactic Database [NED], col. 4-9), redshift (NED,
col. 10), 
absolute magnitude in 
$K$ and $B$ (from McLeod \& Rieke 1995, col. 11 and 12), and Sy host
galaxy type
(from HST observations of Malkan et al. 1998, col. 13)}
\end{table}
}

\begin{table}[h]
\begin{center}
\begin{tabular}{lcrrlrrlc}
\\
\\
\hline
\hline
Galaxy &   Type &  \multicolumn{3}{c}{RA} & \multicolumn{3}{c}{Dec} & z \\
\hline
NGC 1093 & .SX.2?.  & 02&48&16.11&  +34&25&11.4    &  0.018   \\ 
UGC 3247 & .S?....  & 05&06&38.1&    +08&40&24    &  0.011   \\ 
UGC 3407 & .S..1..  & 06&09&08.081& +42&05&05.85    &  0.012   \\
UGC 3463 & .SXS4..  & 06&26&55.18&  +59&04&41.1    &  0.009   \\
UGC 3536 & .L.....  & 06&46&03.74&  +29&20&52.7    &  0.016  \\
UGC 3576 & .SBS3..  & 06&53&06.96&  +50&02&02.0    &  0.020   \\
UGC 3592 & RSBS1..  & 06&55&13.83&  +40&20&16.2    &  0.044   \\
NGC 2347 & PSAR3*.  & 07&16&04.35&  +64&42&36.4    &  0.015   \\
UGC 3789 & RSAR2..  & 07&19&31.53&  +59&21&21.1    &  0.011 \\
NGC 2365 & .SX.1..  & 07&22&22.33&  +22&05&00.6    &  0.008   \\
UGC 3850 & PSXS1..  & 07&28&14.97&  +63&15&21.0    &  0.016 \\
NGC 2431 & PSBS1*.  & 07&45&13.393& +53&04&30.39    &  0.019   \\
NGC 2460 & .SAS1..  & 07&56&52.77&  +60&21&00.0    &  0.005   \\
NGC 2487 & .SB.3..  & 07&58&20.10&  +25&08&59.0    &  0.008   \\
NGC 2599 & .SA.1..  & 08&32&11.20&  +22&33&37.4    &  0.016   \\
NGC 2855 & RSAT0..  & 09&21&27.28&  -11&54&35.5    &  0.006   \\
NGC 3066 & PSXS4P.  & 10&02&11.071& +72&07&31.43    &  0.007   \\
NGC 3188 & RSBR2..  & 10&19&42.75&  +57&25&24.5    &  0.026   \\
NGC 3455 & PSXT3..  & 10&54&31.37&  +17&17&08.2    &  0.004   \\
NGC 4146 & RSXS2..  & 12&10&18.44&  +26&25&53.7    &  0.022   \\
NGC 4369 & RSAT1..  & 12&24&36.18&  +39&22&58.2    &  0.003   \\
NGC 4956 & .L.....  & 13&05&00.91&  +35&10&40.6    &  0.016   \\
NGC 4966 & .S.....  & 13&06&17.20&  +29&03&47.2    &  0.023   \\
NGC 5434 & .SA.5..  & 14&03&23.11&  +09&26&51.5    &  0.019   \\
NGC 5534 & PSXS2P*  & 14&17&40.57&  -07&25&01.1    &  0.009   \\
NGC 5832 & .SBT3\$. & 14&57&45.34&  +71&40&55.3    &  0.001    \\
NGC 5869 & .L..0*.  & 15&09&49.29&  +00&28&13.2    &  0.007   \\
UGC 9965 & .SAT5..  & 15&40&06.79&  +20&40&52.7    &  0.014   \\
NGC 5992 & .S.....  & 15&44&21.457& +41&05&07.85    &  0.032   \\
NGC 6085 & .S..1..  & 16&12&35.225& +29&21&54.16    &  0.034  \\
NGC 6278 & .L.....  & 17&00&50.13&  +23&00&40.7    &  0.009   \\
NGC 6504 & .S.....  & 17&56&05.836& +33&12&29.99    &  0.016   \\
NGC 6635 & .L...P*  & 18&27&37.04&  +14&49&06.6    &  0.017   \\
NGC 6922 & .SAT5P*  & 20&29&52.90&  -02&11&23.8    &  0.019   \\
\hline
\end{tabular}
\end{center}
\caption{Properties of our control sample galaxies. Columns are as in 
Table 1: name (col. 1), morphological classification 
(2), RA and dec (3--8), and
redshift (9)}
\end{table}

\begin{table}[h]
\begin{center}
\begin{tabular}{llccc}
\\
\\
\hline
\hline
Galaxy & Obs. Run & \multicolumn{3}{c}{Seeing ($''$)}\\
~ & ~ & J & H & K \\
\hline
Mrk 334  & Sep 95/WHT  	  &  1.04  &	 1.12 &   0.74 \\
Mrk 335  & Sep 95/WHT  	  &  1.27  &	 1.09 &   0.75 \\
UGC 524  & Sep 95/WHT  	  &  0.72  &	 0.91 &   0.80 \\
I Zw 1   & Sep 95/WHT  	  &  0.77  &	 0.75 &   0.79 \\
Mrk 993  & Sep 95/WHT  	  &  0.81  &	 0.74 &   0.69 \\
Mrk 573  & Sep 95/WHT  	  &  0.81  &	 0.77 &   0.80 \\
UGC 1395 & Sep 95/WHT  	  &  0.86  &	 0.74 &   0.74 \\
Mrk 590  & Sep 95/WHT  	  &  0.99  &	 0.78 &   0.87 \\
NGC 1068 & Sep 95/WHT  	  &  0.95  &	 0.82 &   0.69 \\
NGC 1144 & Sep 95/WHT  	  &  0.89  &	 0.81 &   0.77 \\
Mrk 1243 & Feb 96/CFHT 	  &  0.71  &	 0.75 &   0.76 \\
NGC 3227 & Apr 96/WHT 	  &  0.73  &	 0.68 &   0.62 \\
NGC 3362 & Apr 96/WHT 	  &  0.59  &	 0.69 &   0.75 \\
UGC 6100 & Apr 96/WHT 	  &  0.99  &	 0.94 &   0.91 \\
NGC 3516 & Feb 96/CFHT 	  &  0.71  &	 0.73 &   0.72 \\
Mrk 744  & Feb 96/CFHT 	  &  0.98  &	 1.09 &   0.81 \\
NGC 3982 & Feb 96/CFHT 	  &  0.73  &	 0.73 &   0.73 \\
NGC 4051 & Feb 96/CFHT 	  &  0.73  &	 0.73 &   0.73 \\
NGC 4151 & Apr 96/WHT 	  &  0.62  &	 0.61 &   0.61 \\
NGC 4235 & Feb 96/CFHT 	  &  0.99  &	 1.27 &   0.95 \\
Mrk 766  & Feb 96/CFHT 	  &  1.04  &	 0.93 &   1.00 \\
Mrk 205  & Apr 96/WHT 	  &  0.67  &	 0.69 &   0.73 \\
NGC 4388 & Feb 96/CFHT 	  &  0.77  &	 0.92 &   0.85 \\
Mrk 231  & Feb 96/CFHT 	  &  0.99  &	 0.99 &   1.03 \\
NGC 5033 & Feb 96/CFHT 	  &  1.00  &	 0.75 &   0.87 \\
Mrk 789  & Apr 96/WHT 	  &  0.71  &	 0.69 &   0.59 \\
UGC 8621 & Apr 96/WHT 	  &  0.77  &	 0.76 &   0.87 \\
NGC 5252 & Apr 96/WHT 	  &  0.64  &	 0.71 &   0.60 \\
Mrk 266  & Apr 96/WHT 	  &  0.70  &	 0.65 &   0.67 \\
Mrk 270  & Apr 96/WHT 	  &  0.70  &	 0.65 &   0.67 \\
NGC 5273 & Apr 96/WHT 	  &  0.66  &	 0.83 &   0.70 \\
Mrk 461  & Feb 96/CFHT 	  &  0.90  &	 0.97 &   0.71 \\
Mrk 279  & Apr 96/WHT 	  &  0.73  &	 0.69 &   0.84 \\
NGC 5548 & Apr 96/WHT 	  &  0.66  &	 0.62 &   0.66 \\
NGC 5674 & Apr 96/WHT 	  &  0.74  &	 0.78 &   0.74 \\
Mrk 817  & Sep 95/WHT  	  &  0.89  &	 0.93 &   0.75 \\
Mrk 686  & Sep 95/WHT  	  &  0.76  &	 0.75 &   0.78 \\
Mrk 841  & Sep 95/WHT  	  &  0.92  &	 0.87 &   0.77 \\
NGC 5929 & Sep 95/WHT  	  &  1.43  &	 1.11 &   0.74 \\
NGC 5940 & Sep 95/WHT  	  &  0.83  &	 0.74 &   0.68 \\
NGC 6104 & Sep 95/WHT  	  &  0.75  &	 0.72 &   0.62 \\
UGC 12138& Sep 95/WHT  	  &  0.65  &	 0.70 &   0.68 \\
NGC 7469 & Sep 95/WHT  	  &  0.73  &	 0.75 &   0.65 \\
Mrk 530  & Sep 95/WHT  	  &  0.67  &	 0.70 &   0.73 \\
Mrk 533  & Sep 95/WHT  	  &  0.75  &	 0.70 &   0.71 \\
NGC 7682 & Sep 95/WHT  	  &  1.03  &	 10.2 &   0.98 \\
\hline
\end{tabular}
\end{center}
\caption{Dates of observing run, telescope used, and seeing values 
in each band, for the Sy sample}
\end{table}

\begin{table}[h]
\begin{center}
\begin{tabular}{llc}
\\
\\
\hline
\hline
Galaxy & Obs. Run & Seeing ($''$)\\
~ & ~ &  K  \\
\hline
NGC 1093 &Feb 96/CFHT & 0.94  \\
UGC 3247 &Feb 96/CFHT & 0.59  \\
UGC 3407 &Feb 96/CFHT & 0.79  \\
UGC 3463 &Feb 96/CFHT & 0.85  \\
UGC 3536 &Feb 96/CFHT & 0.66  \\
UGC 3576 &Feb 96/CFHT & 0.82  \\
UGC 3592 &Feb 96/CFHT & 0.81  \\
NGC 2347 &Feb 96/CFHT & 0.86  \\
UGC 3789 &Feb 96/CFHT & 0.81  \\
NGC 2365 &Feb 96/CFHT & 0.74  \\
UGC 3850 &Feb 96/CFHT & 1.01  \\
NGC 2431 &Feb 96/CFHT & 1.00  \\
NGC 2460 &Feb 96/CFHT & 1.04  \\
NGC 2487 &Feb 96/CFHT & 0.85  \\
NGC 2599 &Feb 96/CFHT & 0.90  \\
NGC 2855 &Feb 96/CFHT & 0.80  \\
NGC 3066 &Feb 96/CFHT & 0.88  \\
NGC 3188 &Feb 96/CFHT & 1.01  \\
NGC 3455 &Sep 95/WHT  &  --   \\
NGC 4146 &Sep 95/WHT  & 0.71  \\
NGC 4369 &Sep 95/WHT  &  --   \\
NGC 4956 &Feb 96/CFHT & 0.64  \\
NGC 4966 &Feb 96/CFHT & 0.60  \\
NGC 5434 &Sep 95/WHT  &  --   \\
NGC 5534 &Feb 96/CFHT & 0.90  \\
NGC 5832 &Sep 95/WHT  & 0.80  \\
NGC 5869 &Sep 95/WHT  & 0.69  \\
UGC 9965 &Sep 95/WHT  & 0.80  \\
NGC 5992 &Sep 95/WHT  & 0.75  \\
NGC 6085 &Sep 95/WHT  & 0.64  \\
NGC 6278 &Sep 95/WHT  & 0.78  \\
NGC 6504 &Jun 94/UKIRT & 0.80  \\
NGC 6635 &Sep 95/WHT  & 0.69  \\
NGC 6922 &Sep 95/WHT  & 0.69  \\
\hline
\end{tabular}
\end{center}
\caption{As Table 3, for the control sample}
\end{table}

\end{document}